\documentclass[prl,twocolumn,showpacs,floatfix]{revtex4-1}
\usepackage{graphics,epsfig,amsfonts,amssymb,amsmath,soul,bm,color}
\begin{document}

\title{Domain walls within domain walls in wide ferromagnetic strips}

\author{Touko Herranen}
\email{touko.herranen@aalto.fi}
\author{Lasse Laurson}

\affiliation{COMP Centre of Excellence and Helsinki Institute of Physics,
Department of Applied Physics, Aalto University, P.O.Box 11100, 
FI-00076 Aalto, Espoo, Finland.}

\begin{abstract}
We carry out large-scale micromagnetic simulations which demonstrate that
due to topological constraints, internal domain walls (Bloch lines) within 
extended domain walls are more robust than domain walls in nanowires. 
Thus, the possibility of spintronics applications based on their 
motion channeled along domain walls emerges. Internal domain walls are 
nucleated within domain walls in perpendicularly magnetized media concurrent 
with a Walker breakdown-like abrupt reduction of the domain wall velocity 
above a threshold driving force, and may also be generated within pinned, 
localized domain walls. We observe fast field and current driven internal
domain wall dynamics without a Walker breakdown along pinned domain 
walls, originating from topological protection of the internal domain wall 
structure due to the surrounding out-of-plane domains.
\end{abstract}
\pacs{75.60.Ch, 75.78.Fg, 75.78.Cd}
\maketitle

During recent years, a lot of effort has been devoted to understand properties 
of magnetic domain walls (DWs) and their dynamics. A major driving force behind 
these studies is the emergence of next generation ICT components based on DWs, 
such as memory devices \cite{parkin2008magnetic,parkin2015memory} and logic gates 
\cite{allwood2005magnetic}. At the same time, magnetic DWs constitute
a suitable playground to study several key fundamental physics ideas and concepts, 
ranging from topology \cite{PhysRevLett.95.197204} to non-equilibrium critical 
phenomena \cite{zapperi1998dynamics,papanikolaou2011universality}.

In general, DWs may have various internal degrees of freedom
\cite{PhysRevB.80.054413} which are essential for their magnetic field or 
spin-polarized current driven dynamics, and have recently been shown to be useful
e.g. for channeling spin waves along extended DWs \cite{garcia2015narrow}. In narrow 
ferromagnetic (nano)strips [both in soft Permalloy strips, and strips with a 
high perpendicular magnetic anisotropy (PMA)], the Walker breakdown 
\cite{schryer1974motion}, or the onset of precession 
of the DW internal magnetization ${\bf m}^\text{DW}$, leads to 
an abrupt decrease of the DW propagation velocity above the Walker field 
$B_\text{W}$ or current density $j_\text{W}$
\cite{martinez2012stochastic,martinez2012static,thiaville2006domain}. 
Such behaviour is captured also by simple one-dimensional (1D) models
\cite{thiaville2006domain} of DWs in nanowires/strips.

\begin{figure}[hb!]
\leavevmode
\includegraphics[trim=0cm 0.75cm 0cm 0.2cm, clip=true,width=0.9\columnwidth]
{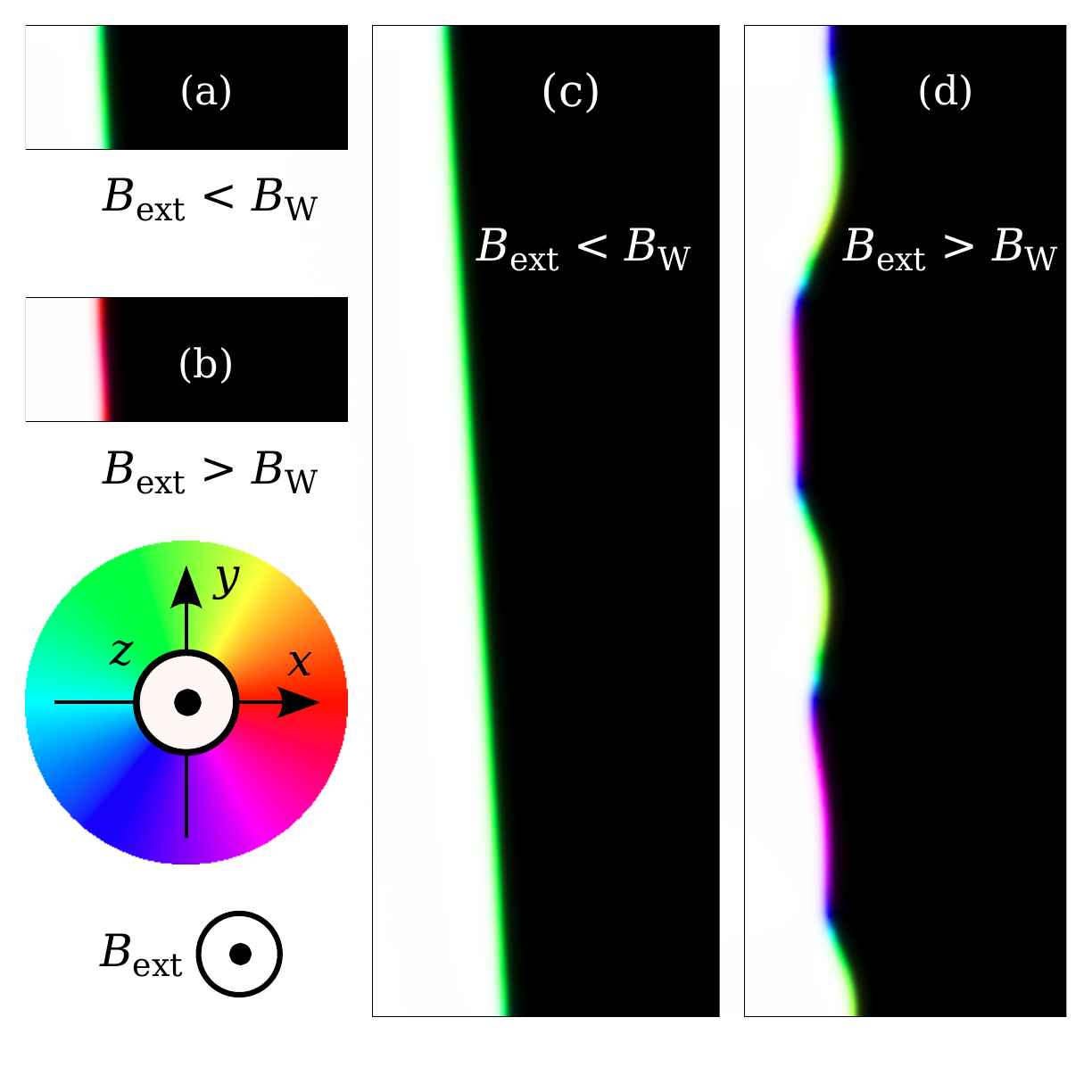}
\caption{(color online) Snapshots of field-driven DWs in 
perfect CoPtCr strips with $B_\text{ext}$ along $z$. (a) For a narrow strip 
($L_y = $ 150 nm), the DW is a
Bloch DW with ${\bf m}^\text{DW}$ roughly along $y$ 
for $B_\text{ext} < B_\text{W}$, while (b) for $B_\text{ext} > B_\text{W}$, 
${\bf m}^\text{DW}$ oscillates between
spatially almost uniform configurations. (c) For a wider 
strip ($L_y = $ 1.2 $\mathrm{\mu}$m), ${\bf m}^\text{DW}$ is uniform 
for $B_\text{ext} < B_\text{W}$, but (d) breaks into a dynamic
pattern of in-plane domains separated by IDWs for $B_\text{ext} > B_\text{W}$.}
\label{fig:no_wb-wb}
\end{figure}

In wider PMA strips with longer DWs (such that the 1D models are no longer 
applicable), one may expect that a Walker breakdown -like
abrupt reduction of the DW propagation velocity still takes place, but that the 
related excitation(s) of ${\bf m}^\text{DW}$ can no longer be spatially uniform 
\cite{zebrowski1981dynamic}. Possible origins for the symmetry breaking leading
to incoherent precession of ${\bf m}^\text{DW}$ in different parts of the DW could
be edge effects, and/or quenched disorder, interacting with 
the DW \cite{leliaert2015thermal,leliaert2014current,
PhysRevB.86.144415,martinez2012stochastic,nakatani2003faster,min2010effects,
PhysRevB.89.064419,jiang2010enhanced}; these may include dislocations, 
precipitates, grain boundaries, thickness fluctuations of the strip, 
etc. Here we explore the dynamics of extended 
DWs in wide CoPtCr PMA strips, with a Bloch wall equilibrium structure, using 
large-scale micromagnetic simulations with and without quenched disorder. 
We show that in wide enough strips and for driving forces exceeding a threshold
value ($B_\text{W}$ or $j_\text{W}$),
the internal degrees of freedom of the DW are indeed excited inhomogeneously 
along the DW. These non-uniformities can be well-described as a set of internal 
DWs (IDWs), similar to Bloch lines observed e.g. in the context of bubble domains 
in garnet films \cite{volkov2008domain,konishi1983new,malozemoff1972effect,
slonczewski1974theory,malozemoff1979magnetic}, separating in-plane domains 
within the main DW, see Fig. \ref{fig:no_wb-wb} (d) for an example snapshot. 
Such IDWs, which resemble transverse DWs \cite{hertel2015analytic} found 
in narrow nanostrips with in-plane magnetization \cite{mcmichael1997head,
nakatani2005head, estevez2015head}, can also be nucleated within DWs pinned 
by notches, and subsequently driven along the localized 
main DW by magnetic fields or spin-polarized currents. We compute the driving
force dependence of the velocity of the various IDW types observed, and 
find that they do not experience a Walker breakdown, due to topological 
constraints originating from the out-of-plane domains surrounding the IDWs.
Thus, the possibility to use extended DWs as channels for fast IDW 
propagation emerges, with potential applications in spintronics. 

The micromagnetic simulations are performed using the GPU-accelerated micromagnetic 
code MuMax3 \cite{mumax_git,mumax2011,mumax2014}, solving  
the Landau-Lifshitz-Gilbert equation \cite{gilbert2004phenomenological,brown1963micromagnetics},
\begin{equation}
\label{eq:llg}
\partial {\bf m}/\partial t =
\gamma {\bf H}_\text{eff} \times {\bf m} + \alpha {\bf m} \times
\partial {\bf m}/\partial t,
\end{equation}
where ${\bf m}$ is the magnetization, $\gamma$ the 
gyromagnetic ratio, and ${\bf H}_\text{eff}$ the effective
field, with contributions due to exchange, Zeeman, and demagnetizing
energies. We consider CoPtCr strips of thickness $L_z=$ 20 nm  \cite{kubota1998}, and 
widths $L_y$ ranging from 150 nm to 3 $\mu$m, with the saturation magnetization 
$M_\text{s}$ = 3 $\cdot$ $10^5$ A/m, exchange constant $A$ = $10^{-11}$ J/m, and 
the damping parameter $\alpha$ = 0.2. The first order uniaxial anisotropy 
constant $K_\text{u} = 2 \cdot 10^5$ $ \mathrm{J/m^3}$, 
and an out-of-plane easy axis is considered, to take into account the PMA 
nature of the strip \cite{weller2000}. The discretization cells have 
dimensions of $\Delta_x$ = $\Delta_y$ = 3 nm, and $\Delta_z$ = 20 nm; we 
chose to use only one layer of computational cells in the $z$-direction, 
since a finer discretization in that direction did not change the results.

We start by considering the field-driven dynamics of DWs in perfect
CoPtCr strips, with two out-of-plane domains separated by a Bloch DW with 
the internal magnetization ${\bf m}^\text{DW}=+M_\text{s}\hat{y}$ as an initial 
state. We employ a simulation window of length
$L_x$ = 6 $\mu$m centered around and moving with the DW. When driven with a field 
of magnitude $B_\text{ext} < B_\text{W}(L_y)$ along $z$, we observe a slight, 
$B_\text{ext}$-dependent tilting of the propagating DW, see Figs. \ref{fig:no_wb-wb} (a) 
and (c): as $B_\text{ext} \hat{z}$ rotates ${\bf m}^\text{DW}$ away from the 
$y$ direction, the DW tries to partially align itself with its ${\bf m}^\text{DW}$, 
to minimize stray fields. Notice that this effect arises here without 
the Dzyaloshinskii-Moriya interaction \cite{boulle2013domain}.
For $B_\text{ext} > B_\text{W}$, the DW dynamics depends 
strongly on the strip width: for narrow strips 
[see Fig. \ref{fig:no_wb-wb} (b)], ${\bf m}^\text{DW}$ oscillates between spatially
almost uniform configurations, while for wider strips with longer DWs, ${\bf m}^\text{DW}$
breaks into a dynamic, spatially non-uniform pattern, nucleated from the sample edges; 
snapshots of the magnetization configurations [Fig. 
\ref{fig:no_wb-wb} (d)] reveal that ${\bf m}^\text{DW}$ is broken into
a set of in-plane domains, separated by IDWs (Bloch lines). These patterns are highly
dynamic, with different parts of the DW either ahead or behind the average
DW position. The IDWs move along the narrow (width $\sim 9$ nm) main DW, with 
annihilation and nucleation events of pairs of IDWs taking place 
repeatedly \cite{zvezdin1990dynamic}; an example animation of this complex process 
is provided as Supplemental Material \cite{SM}. 

\begin{figure}[t!]
\leavevmode
\includegraphics[trim=0cm 0cm 0cm 0cm, clip=true,width=0.9\columnwidth]
{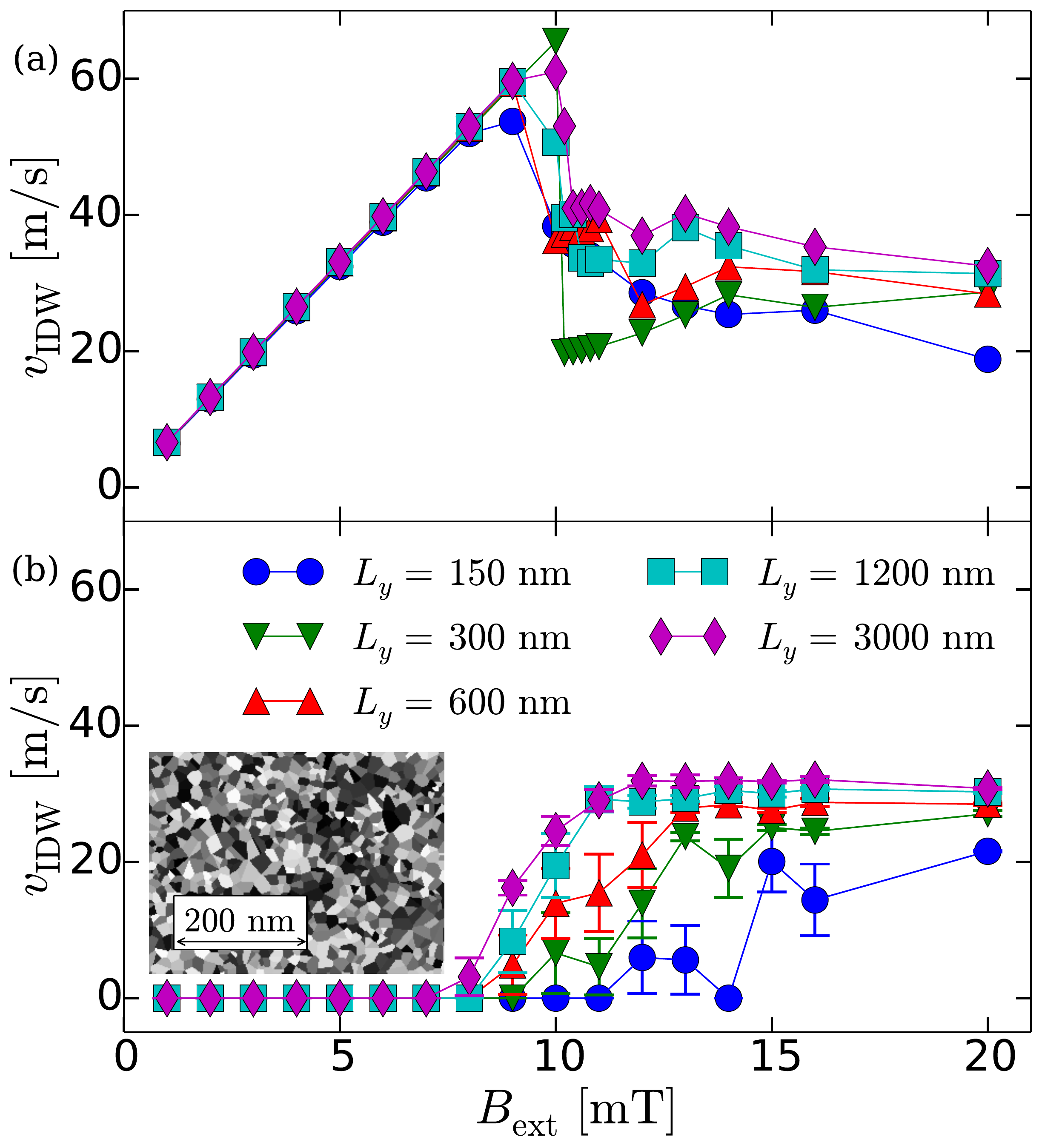}
\caption{(color online) The average DW velocity $v_\text{DW}$ as a function of 
$B_\text{ext}$ in CoPtCr strips of different widths $L_y$, in
(a) perfect strips and (b) strips with a random 7 \% standard deviation 
variation of the anisotropy strength $K_\text{u}$ in each grain; an example of
the grain structure is shown in the inset. Notice the 
strong size effect in the depinning field in (b).}
\label{fig:width_velo}
\end{figure}

Fig. \ref{fig:width_velo} (a) shows the steady state DW velocity $v_\text{DW}$ 
as a function of $B_\text{ext}$ for perfect strips of different widths $L_y$.
For all $L_y$-values considered, $v_\text{DW}$ exhibits 
the same linear dependence on $B_\text{ext}$ for $B_\text{ext} < B_\text{W}(L_y)$,
with $B_\text{W}(L_y)$ in the range 9-10 mT; above $B_\text{W}$, 
$v_\text{DW}$ is sigficantly smaller than its (local) maximum obtained for smaller 
fields (indeed, a reduction of $v_\text{DW}$ by a factor of $\alpha^2$ is expected 
in the precence of many Bloch lines \cite{malozemoff1979magnetic,slonczewski1974theory,
malozemoff1972effect}), and displays a relatively complex depedence on $B_\text{ext}$, 
originating from the incoherent dynamics of ${\bf m}^\text{DW}$ at different parts 
of the DW. Also current-driven DW dynamics (not shown) exhibits a velocity drop and IDWs 
nucleated above $j_\text{W}$ \cite{iwasaki2015current}.

\begin{figure*}[ht!]
\leavevmode
\includegraphics[trim=0cm 0cm 0cm 0cm, clip=true,width=1.8\columnwidth]
{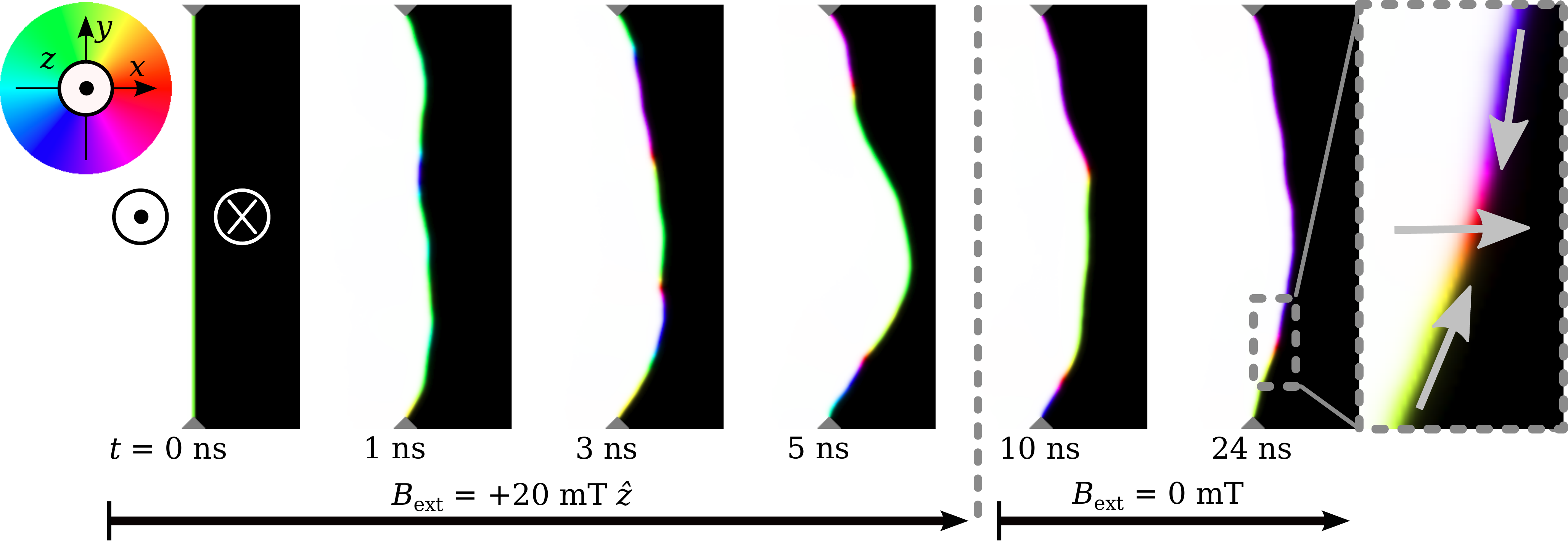}
\caption{(color online)
IDWs may be generated within pinned DWs in disordered CoPtCr strips by field 
pulses. A field pulse of magnitude $B_\text{ext}=20$ mT $> B_\text{W}$ and duration
of 5 ns, followed by relaxation in zero field, results in one pinned IDW, shown 
magnified in the last frame. The grey arrows indicate the direction of the 
in-plane DW magnetization around the IDW.}
\label{fig:generation}
\end{figure*}

In order to account for the effect of quenched disorder, here assumed to originate 
from the polycrystalline structure of the strip \cite{disorder_remark}, we construct 
grains using a Voronoi tessellation \cite{leliaert2015thermal,leliaert2014current}, 
with an average grain size of 11.9 nm \cite{kubota1998}, see the inset of Fig. 
\ref{fig:width_velo}. 
For each grain $i$, we assign a different, Gaussian distributed random anisotropy 
strength $K_{\text{u},i}$, with mean $K_{\text{u}}$ and standard deviation 
$\sigma = 0.07K_{\text{u}}$. Here, we consider strips of a finite length $L_x$ = 6 $\mu$m. 
The DW is initialized at $x$ = −1.5 $\mu$m, with the origin being in
the middle of the sample. The disorder results in a finite, $\sigma$-dependent 
depinning field, as evidenced by 
Fig. \ref{fig:width_velo} (b), where averages over 5 disorder realizations
are presented. Notice the strong size effect in the depinning field, with the 
longer DWs depinning for smaller $B_\text{ext}$. For longer DWs, 
${\bf m}^\text{DW}$ now precesses non-uniformly for all $B_\text{ext}$ 
values with $v_\text{DW}>0$, including those below $B_\text{W}$ of the corresponding 
perfect system; thus, the non-uniform internal degrees of freedom of the DW 
can also be induced by strong enough disorder, in addition to the sample edges.

IDWs can also be created within localized, pinned main DWs; such setups could
be useful to experimentally test our results.
We consider here DWs pinned by triangular notches (with 50 nm long sides, and
a $50 \sqrt{2} \approx 70.7$ nm long base) in strips of width $L_y$ = 1.2 $\mathrm{\mu}$m
and length $L_x$ = 6 $\mathrm{\mu}$m, with disorder $\sigma  = 0.1 K_\text{u}$.
By applying a 5 ns long square field pulse of amplitude $B_\text{ext} = \pm 20$ mT 
(i.e. above $B_\text{W}$) along $z$, the pinned DW bends as its central
part propagates, and ${\bf m}^\text{DW}$ exhibits spatially non-uniform dynamics 
(as  $B_\text{ext}>B_\text{W}$), see Fig. \ref{fig:generation}. After the pulse, the 
curved DW gets pinned by disorder, and may contain long-lived pinned non-uniformities 
in its ${\bf m}^\text{DW}$: stable $180^\circ$ IDWs (Bloch lines) within the pinned 
main DW are formed with a roughly 40 \% success rate (estimated from 
an ensemble of 100 disorder realizations). Also more complex structures are sometimes 
observed, such as $360^\circ$ IDWs. Fig. \ref{fig:generation} shows a typical example 
of the process, with the end result of one head-to-head (H2H) $180^\circ$ IDW, pinned 
by the disorder even after the relaxation time of 19 ns in zero field; the same process 
is also showed in a Supplemental Movie \cite{SM}. Magnification 
of the IDW magnetization (last frame of Fig. \ref{fig:generation}) 
reveals that the IDW resembles transverse DWs in narrow in-plane 
systems (e.g. Permalloy nanostrips) \cite{hertel2015analytic,mcmichael1997head,
nakatani2005head}. In addition to giving rise to
different curvatures of the main DW, the direction ($\pm z$) of the applied field 
pulse affects the polarity of the IDWs created; 
$B_\text{ext} > 0$ tends to lead to IDWs with $m_x^\text{IDW} > 0$ (as e.g. in 
Fig. \ref{fig:generation}), whereas $B_\text{ext} < 0$ 
gives mostly rise to IDWs with $m_x^\text{IDW} < 0$; for both polarities, 
the IDW may have either a H2H or tail-to-tail (T2T) configuration.

\begin{figure}[t!]
\leavevmode
\includegraphics[trim=0cm 0cm 0cm 0cm, clip=true,width=0.9\columnwidth]
{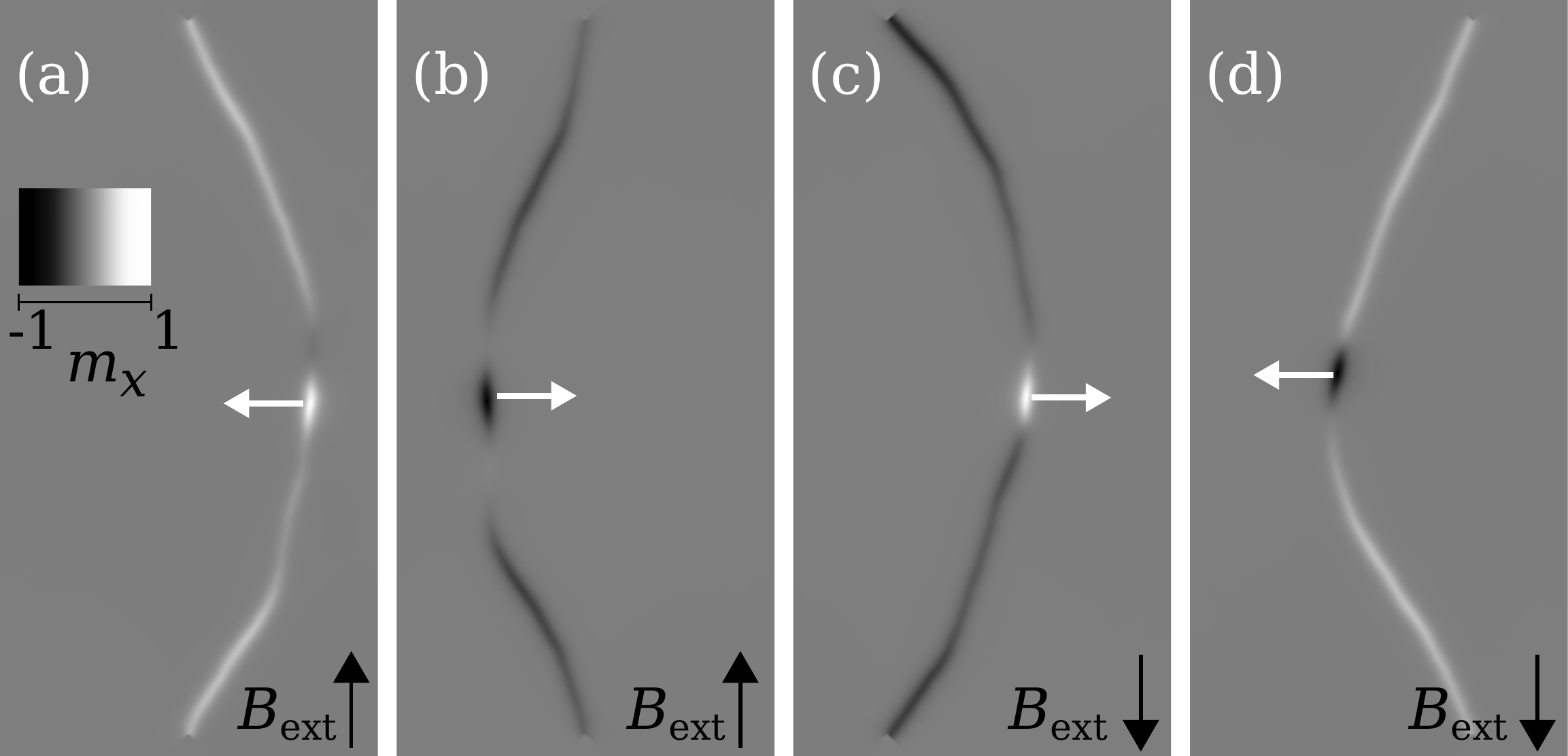}
\caption{Snapshots of IDWs of different structures, driven from
bottom to top by a field $B_\text{ext}$ along the $\pm y$
direction, shown as black arrows. The grayscale indicates the magnitude
of $m_x$. IDWs in (a) and (b) are of the H2H type,
whereas (c) and (d) have T2T configurations. Depending on the
sign of ${\bf m}^\text{IDW}$, $B_\text{ext}$
gives rise to a force acting on the internal DW either towards left or
right, as shown by the white arrows.}
\label{fig:mechanism}
\end{figure}

\begin{figure}[t!]
\leavevmode
\includegraphics[trim=0cm 0cm 0cm 0cm, clip=true,width=0.95\columnwidth]
{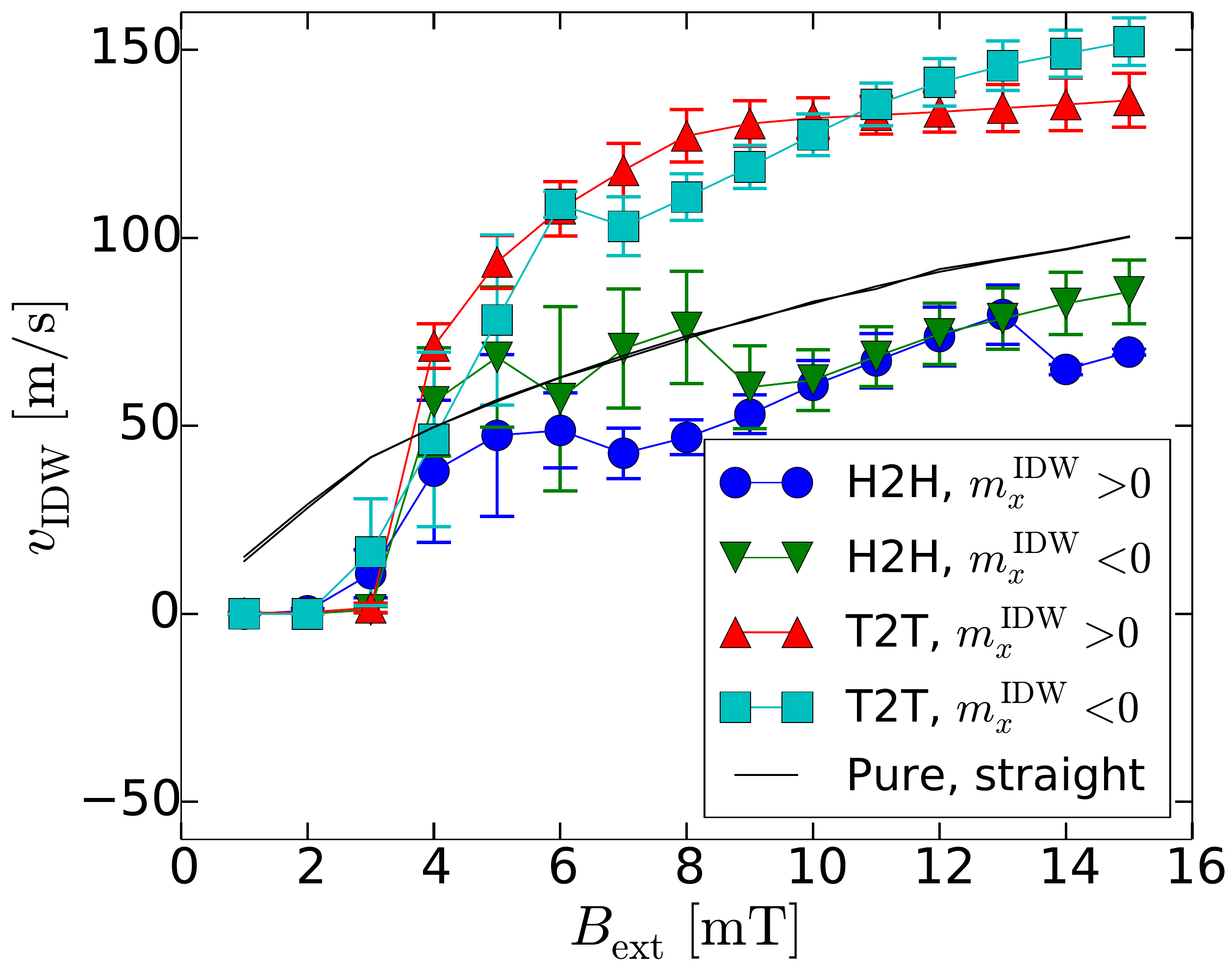}
\caption{(color online) The average (over 4 disorder realizations) IDW 
velocity $v_\text{IDW}$ as a function of $B_\text{ext}$ for the four 
possible IDW configurations. The black lines show the corresponding
data for perfect strips with artificially generated straight main DWs,
each containing an IDW of one of the four different kinds (H2T,T2T,
$m_x^\text{IDW}>0$, $m_x^\text{IDW}<0$).}
\label{fig:velos_dwindw}
\end{figure}

Examples of these four IDW configurations are presented 
in Fig. \ref{fig:mechanism}. Due to the geometry 
shown in Fig. \ref{fig:mechanism}, the IDWs can be driven by applying 
fields along $\pm y$ directions. To study the velocity-field characteristics
of the IDWs, we first move all of them from their random initial positions to 
close to the lower edge of the strip by applying field pulses 
of a small magnitude. Then, a driving field $B_\text{ext}$ is applied
along $+y$ or $-y$ direction, to move the IDW towards the upper edge;  
the sign depends on whether the IDW has a H2H or T2T structure. The 
resulting $v_\text{IDW}(B_\text{ext})$ curves are shown in Fig. \ref{fig:velos_dwindw}, 
separately for the four possible IDW configurations,
revealing that for $B_\text{ext}$ above a small 
depinning field, the T2T IDWs move significantly faster than the
corresponding H2H IDWs. This can be understood by considering
the force due to $B_\text{ext}$ on the IDW (with ${\bf m}^\text{IDW}$ along $\pm x$) 
in directions perpendicular to the main DW, shown with white arrows in 
Fig. \ref{fig:mechanism}: for H2H IDWs, the force direction coincides with 
that due to the line tension of the curved main DW [Fig. \ref{fig:mechanism} 
(a) and (b)]. Thus, during the dynamics of the IDW, part of the energy of 
the driving field is dissipated in partial straightening of the main DW, 
resulting in a lower $v_\text{IDW}$ for H2H IDWs. In contrast, for T2T IDWs, 
the perpendicular force on the IDW due to the driving field points in the opposite
direction from that of the curvature-induced force: thus, less motion of the
main DW takes place during the IDW dynamics, and a larger fraction of the field 
energy is available to move the T2T IDW, resulting in a larger $v_\text{IDW}$.
This is confirmed by movies provided as Supplemental Material \cite{SM}:
the motion of T2T IDWs is noticeably smoother than that of H2H IDWs.
As a further check, we consider also a pure system with an artificially 
generated straight main DW containing an IDW; the resulting $v_\text{IDW}(B_\text{ext})$
curves for the four possible configurations are shown as solid black lines in 
Fig. \ref{fig:velos_dwindw}. All cases exhibit the same $v_\text{IDW}(B_\text{ext})$
behaviour, which is intermediate between those found for curved main DWs
with H2H and T2T IDWs, respectively. Thus, the perpendicular
forces still play a role in energy dissipation, but less than in the curved 
H2H case.

The $v_\text{IDW}(B_\text{ext})$ curves in Fig. \ref{fig:velos_dwindw}
exhibit two noteworthy features: (i) $v_\text{IDW}$ grows sub-linearly with 
$B_\text{ext}$, indicating that an increasing fraction of the field energy
is dissipated in other processes than IDW motion when $B_\text{ext}$ is increased.
(ii) No clear velocity drop, a signature of Walker breakdown, is observable;
this is the case even if we consider relatively large fields up to 15 mT. 
Inspection of the magnetization dynamics of the IDWs (see Supplemental 
Material \cite{SM}) reveals that the direction of ${\bf m}^\text{IDW}$
is indeed preserved during the dynamics. This can be understood to 
follow from the peculiar topology of the system at hand: unlike
nanostrips, the main DW along which the IDW is propagating does
not have free boundaries; these are crucial e.g. for nucleation of antivortices, 
mediating the magnetization reversal of transverse DWs in in-plane nanostrips. 
Instead, here the IDWs are surrounded by the two out-of-plane domains, 
which topologically protect ${\bf m}^\text{IDW}$ from being flipped, leading to the 
absence of Walker breakdown, and consequently to relatively large field-driven IDW 
velocities (up to 150 m/s for T2T IDWs). A similar absence of the Walker breakdown
of topological origin has been reported in ferromagnetic nanotubes \cite{yan2011fast}.

\begin{figure}[t!]
\leavevmode
\includegraphics[trim=0cm 0cm 0cm 0cm, clip=true,width=0.95\columnwidth]
{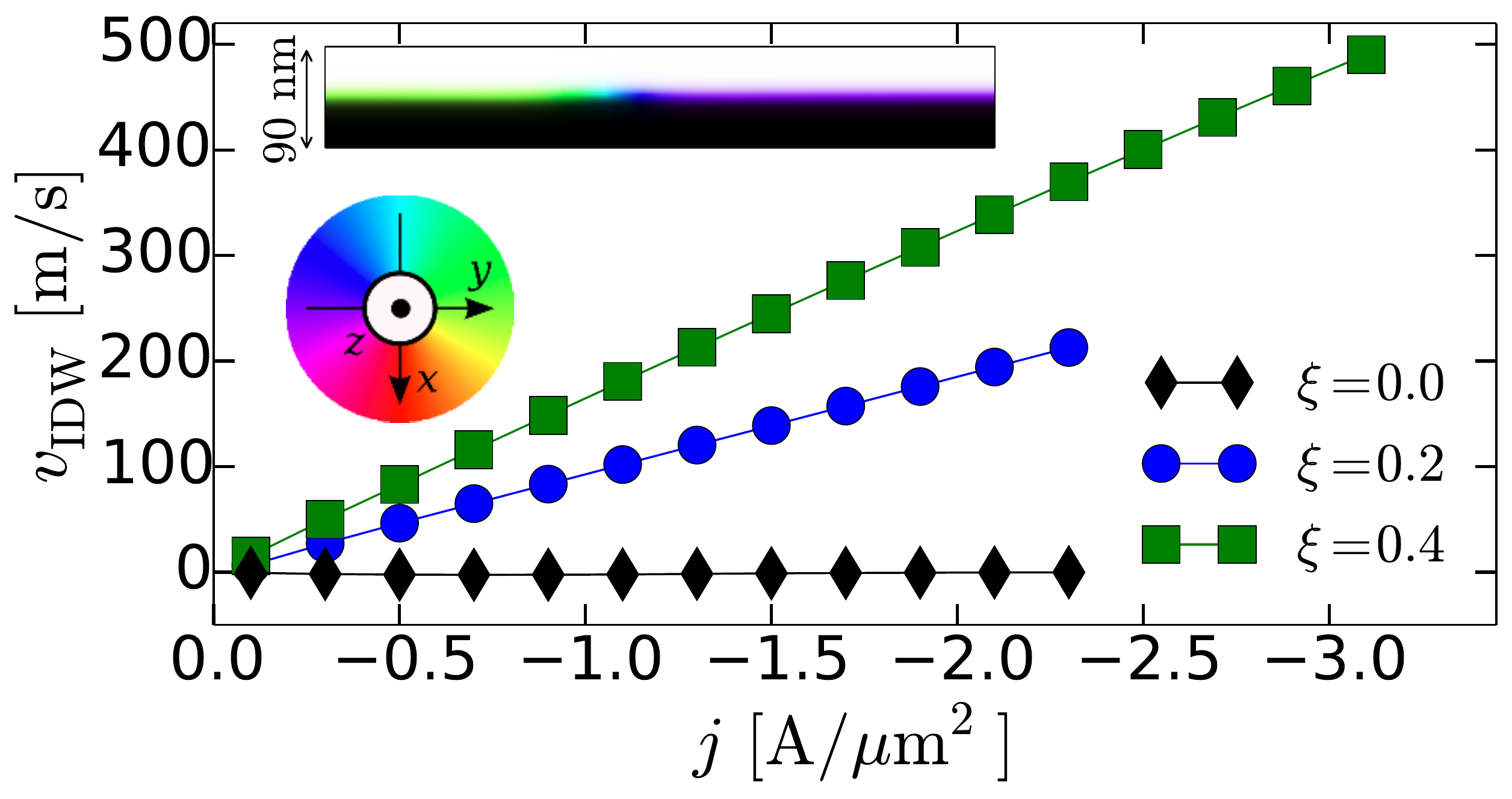}
\caption{(color online) The IDW velocity $v_\text{IDW}$ as a function of the 
current density $j$, for different values of the non-adibatic parameter $\xi$, in 
a narrow strip with the main DW along the long axis of the strip; part of such a 
strip is shown in the inset. For $|j|$-values larger than those shown, the domain 
structure breaks down.}
\label{fig:6}
\end{figure}

Finally, we consider current-driven IDW dynamics in narrow perfect CoPtCr
strips with the main DW located in the middle of the strip and oriented along its 
long axis (inset of Fig. \ref{fig:6}). By extending Eq. (\ref{eq:llg}) with spin-transfer 
torque terms \cite{zhang2004roles}, we observe a simple linear dependence of $v_\text{IDW}$ 
on the current density $j$ for $\xi=\alpha$ and $\xi=2\alpha$ (with $\xi$ the 
non-adiabatic parameter) up to $v_\text{IDW} = 490$ m/s for $\xi=2\alpha$, whereas 
for $\xi=0$, $v_\text{IDW} = 0$ for all $j$ (Fig. \ref{fig:6}). Thus, Walker 
breakdown is absent also in current-driven IDW dynamics, due to the topological 
protection discussed above; this implies also that for $\xi=0$, IDWs remain 
intrinsically pinned \cite{koyama2011observation} for any $j$. 

To conclude, we have shown that for DWs in wide PMA strips, a Walker 
breakdown -like abrupt reduction of the DW propagation velocity is 
concurrent with nucleation of internal in-plane domains separated by internal 
DWs (Bloch lines), resulting in a hierarchical DW structure, with DWs within 
DWs. The IDWs can be driven along the main DW, and the absence of 
Walker breakdown in their dynamics could lead to interesting possibilities 
for spintronics applications where DWs would serve as guides for fast 
IDW propagation.

\begin{acknowledgments}
We thank Mikko J. Alava and Jan Vogel for useful comments.
This work has been supported by the Academy of Finland through its Centres
of Excellence Programme (2012-2017) under project no. 251748, and an Academy
Research Fellowship (LL, project no. 268302). We acknowledge the computational 
resources provided by the Aalto University School of Science ``Science-IT'' 
project, as well as those provided by CSC (Finland).
\end{acknowledgments}

\bibliography{bibl}

\begin{thebibliography}{44}
\bibitem [{dis()}]{disorder_remark}%
  \BibitemOpen
  \href@noop {} {}\bibinfo {note} {We expect other sources of disorder, leading
  to a random energy landscape interacting with the DW, to give rise to similar
  results.}\BibitemShut {Stop}%
\end{thebibliography}

\end{document}